\documentstyle[aaspp4]{article}
\newcommand\etal{{\it et al.\/}\ }
\begin{document}
\title{Asymmetry in Isolated, Morphologically Normal Sa Galaxies}

\author{David A. Kornreich\altaffilmark{1}}
\affil{Center for Radiophysics and Space Research} 
\affil{Cornell University Space Sciences Building,
Ithaca NY 14853}
\affil{kornreic@astro.cornell.edu}
\author{Martha P. Haynes}  
\affil{Center for Radiophysics and Space
Research and National Astronomy and Ionosphere Center\altaffilmark{2}}
\affil{Cornell University Space Sciences Building,
Ithaca NY 14853}
\affil{haynes@astro.cornell.edu}
\author{Katherine P. Jore}
\affil{Department of Physics and Astronomy, University of Wisconsin at Stevens Point, Stevens Point WI 54481}
\affil{kjore@uwsp.edu}
\and
\author{R. V. E. Lovelace}
\affil{Department of Astronomy} 
\affil{Cornell University Space Sciences Building,
Ithaca NY 14853}
\affil{rvl1@cornell.edu}

\altaffiltext{1}{NASA Space Grant Graduate Fellow}
\altaffiltext{2}{The National Astronomy and Ionosphere Center is
operated by Cornell University under a cooperative agreement with the
National Science Foundation.}

\begin{abstract}

We have examined the morphological and dynamical \ion{H}{1} symmetry
properties of a sample of moderately inclined Sa galaxies which are
classified as morphologically normal. The sample galaxies  were
known {\it a priori} to exhibit kinematic peculiarities ranging from warps to
independent, wholly decoupled disks, and are possibly the remnants of
minor mergers. We compare the asymmetry of the rotation curves to
global kinematic asymmetry, and find a relationship between rotation
curve asymmetry and the kinematic $n=2$ mode. We have also examined
the kinematics of these galaxies following the discussion of Briggs
(1990) and find that the warps observed in the \ion{H}{1} disks of these
galaxies deviate significantly from the simple rules for warps that
commonly apply.

\end{abstract}


\section{Introduction}

Minor mergers are a prime candidate for the excitation of asymmetries
in galaxies, both kinematic and morphological. As such, it is
important to investigate the symmetry properties of galaxies believed
to be the remnants of minor mergers when seeking to understand the
nature of the resulting perturbations.

As part of a study to investigate the homogeneity and dark matter
content of the Sa galaxy class, Jore (1997)\markcite{KJPhD}
investigated the dynamics of gas and stars in a sample of 20 nearby,
morphologically normal, isolated Sa galaxies. For nine gas--rich objects,
\ion{H}{1} synthesis maps were obtained with the Very Large
Array (VLA)\footnote{The VLA is a part of the
National Radio Astronomy Observatory, which is a facility of the
National Science Foundation operated under a cooperative
agreement by Associated Universities, Inc.}.

Although Jore's sample was selected to be isolated and morphologically 
normal, several cases of peculiar kinematics were discovered. Jore
\etal (1996; hereafter Paper~I) discussed NGC~4138, an isolated Sa
containing coplanar counter--rotating stellar disks embedded in an
extensive \ion{H}{1} disk. Haynes \etal (2000; hereafter Paper~II)
discuss optical photometry and spectroscopy obtained for four
galaxies: NGC~3626, NGC~3900, NGC~4772, and NGC~5854, and \ion{H}{1}
synthesis line imaging for the first three of these, all of which exhibit
some form of counter--rotation or kinematically distinct ionized gas
components. Jore \etal (2000; hereafter Paper~III) discuss optical and
\ion{H}{1} synthesis data obtained for the remaining VLA targets:
NGC~1169, NGC~3623, NGC~4866, NGC~5377, and NGC~5448. Of these
galaxies, too, three (NGC~3623, NGC~4866, and NGC~5377) exhibit
indications of kinematic decoupling in their optical
spectra. NGC~5448, although kinematically regular as traced optically,
exhibits an inner disk  that is non--coplanar with the outer disk as defined
beyond half the optical radius, $0.5R_{25}$.

The galaxies studied in Papers I---III contain moderate \ion{H}{1}
masses spread over a large area, with $R_{HI}/R_{25}>2$, so that the
neutral gas surface density is very low. The only exception, NGC~3623, 
is gas poor with a shrunken \ion{H}{1} disk. It also is the only
galaxy with nearby major companions, and is in fact a member of the
Leo Triplet (Haynes \etal 1979\markcite{H79}). 

The details of the observations and data analysis leading to the 
conclusions of kinematic decoupling in these galaxies is discussed 
exclusively in Papers~I---III. In this work, we offer no further 
evidence for their existence. Here, instead, our intent is not to 
produce detailed kinematic models of these targets, but rather to 
classify the kinematic and morphological asymmetry exhibited by the 
\ion{H}{1} disks and to compare their global \ion{H}{1} rotation 
curve properties to sets known to be dynamically normal, and thus to 
demonstrate that decoupled components as observed in optical spectra 
occur in galaxies with abnormal warps and global rotation curves in 
the \ion{H}{1}.

The most commonly accepted explanation of the occurrence of
kinematically distinct components 
is that of minor mergers (see, for example, Thakar \etal
1997\markcite{thakar}), which have been proposed to account for the
counter--rotating disks of NGC~7217 (Merrifield \& Kujiken
1994\markcite{MK}; Buta \etal 1995\markcite{buta}), NGC~3593 (Bertola
\etal 1996\markcite{bert}; Corsini \etal 1998\markcite{corsini}),
NGC~3626 (Ciri \etal 1995\markcite{ciri}), and NGC~4138 (Paper
I). Thakar \etal have been able to produce by numerical simulation
morphologically symmetric counter--rotating disks via both gradual
infall and minor mergers. The main drawback is that minor mergers tend
to heat the disk significantly. Conversely, Zaritsky \& Rix
(1997)\markcite{rz97} observe morphological lopsidedness and interpret
their derived statistic that 30\% of field disks are asymmetric as
evidence of ongoing mass infall at a rate of about 2\% of the disk
mass per Gyr. Quinn \etal (1993)\markcite{q} have also numerically
studied the effects of minor mergers, and predict that mergers with
satellite galaxies generally result in a heating of the entire disk
around all three axes, as well as the generation of a long--lived warp
in disk inclination of order $15^\circ$. The numerical work of
Hernquist \& Mihos (1995)\markcite{mh95} studied similar mergers and
characterized their dynamics, finding significant excitation of $n=2$
spiral modes at a level of approximately 20\% of the total observed
power in all Fourier components and of the $n=1$ and $n=3$ modes at
lesser levels.

The deviations from morphological and kinematic axisymmetry can be
quantified in a number of ways. Zaritsky \& Rix quantify morphological
asymmetry by dividing the galaxy into a set of concentric rings and then
measuring the Fourier strengths of the flux through each ring. This
method has the advantage of identifying the type and strength of the
asymmetry as a function of radius, but cannot be used with small,
distant objects due to data scarcity. More widely applicable methods
are those which consider the galaxy as a whole; such methods miss any
radial dependence, but do not suffer as strongly in small data sets
(because they integrate large portions of the image) and are quick and
dirty ways of identifying nonaxisymmetries. Conselice
(1997)\markcite{c97}, for instance, suggests rotating a galaxy image
through 180$^\circ$ (or some other convenient angle) and subtracting
the resultant image from the original. The residuals are taken as the
asymmetry measure. Kornreich \etal (1998\markcite{khl}; hereafter KHL)
propose dividing the galaxy into a number (usually 8) of sectors
centered on the optical center of light of the galaxy and integrating
the flux in each sector. The greatest normalized difference between
sectors is taken as the asymmetry measure.

Deviations from flat, circular motion in galaxies have been quantified
in \ion{H}{1} synthesis data by Kornreich \etal (2000\markcite{HKLvZ};
hereafter KHLvZ). Since flat, circular motion implies a constant
kinematic position angle as a function of radius, kinematic asymmetry
has been quantified as average changes in position angle, either as a
function of radius or between approaching and receding sides in model
velocity fields. Kinematic nonaxisymmetry can also manifest as
differences in the rotation curves of the approaching and receding
sides, and the normalized integrated rotation curve differences are
also taken as quantifiers of non-axisymmetry (Schoenmakers \etal
1997\markcite{SFdZ}, hereafter SFdZ).

Here our goal is to identify and quantify the axisymmetry of the
dynamics of the nine \ion{H}{1}--rich galaxies in the sample of Jore (1997)
mapped with the VLA in terms of the asymmetry
parameters of KHLvZ. The more inclined aspect at which these galaxies
are viewed, with respect to the sample of KHLvZ, allows the
dynamical parameters to be determined with much greater precision. We
will also compare the dynamics of this sample with that of the sample of
Briggs (1990)\markcite{briggs}, a sample of warped but otherwise
kinematically normal galaxies, to determine whether the kinematically
peculiar sample follows the same pattern of warps detected there.

\section{The Sa Galaxy Sample}

The nine objects selected for this analysis are a subset
of the sample of isolated, morphologically normal Sa galaxies
originally selected by Jore (1997)\markcite{KJPhD} in her study
of Sa kinematics and dynamics. The subset discussed here
and summarized in Table~\ref{Kornreich:tab1} were selected
because they are relatively HI--rich, making \ion{H}{1} synthesis observations
with the VLA feasible. The columns of Table~\ref{Kornreich:tab1}
represent:

\begin{deluxetable}{lllll}
\tablenum{1}
\tablecaption{Overview of the Sa Sample}
\tablehead{
\colhead{Galaxy}&
\colhead{Type}&
\colhead{$D_{25} \times d_{25}$}&
\colhead{Peculiar Optical Kinematics}&
\colhead{Ref.}
\\[.2ex]
\colhead{}&
\colhead{}&
\colhead{(arcmin)}&
\colhead{(except where otherwise noted)}&
\colhead{}
}
\startdata
NGC~1169 &	SB(r)a I & 4.17$\times$2.82&	Strong \ion{H}{1} warping&		a \cr	
NGC~3623 &	S(s)a II & 9.77$\times$2.88&	Counter--Rotating Gas Core&		a \cr
NGC~3626 &	Sa	& 2.69$\times$1.95&	Extended Counter--Rotating Gas&		b \cr
NGC~3900 &	S(r)a	& 3.16$\times$1.70&	Central High--Velocity Gas? Streaming?& b \cr
NGC~4138 &	S(r)a pec&2.57$\times$1.70&	Extended Counter--Rotating Gas, Stars&  c \cr
NGC~4772 &	Sa:	& 3.39$\times$1.70&	Counter--Rotating Gas Core&		b \cr
NGC~4866 &	Sa	& 6.31$\times$1.35&	Nonrotating Gas&			a \cr
NGC~5377 &	SBa	& 3.72$\times$2.09&	Counter--Rotating Gas?&			a \cr
NGC~5448 &	S(s)a   & 3.98$\times$1.78&	Twisting of Inner \ion{H}{1} Position Angles&	a \cr

\enddata
\tablenotetext{a}{Jore \etal (2000)}
\tablenotetext{b}{Haynes \etal (2000)}
\tablenotetext{c}{Jore \etal (1996)}
\label{Kornreich:tab1}
\end{deluxetable}

\noindent (1) The NGC designation of the target galaxy;

\noindent (2) The morphological classification as given in the 
{\it Revised Shapley Ames Catalog} (Sandage \& Tammann 1987)\markcite{RSA}.

\noindent (3) The optical major and minor diameters
at 25~mag~arcsec$^{-2}$, taken from 
the {\it Third Reference Catalog of Bright Galaxies}
(de Vaucouleurs \etal 1991\markcite{RC3}; RC3).

\noindent (4) Type of distinct kinematics  detected in the optical
spectroscopy by Papers~I---III, except for NGC~1169 and NGC~5448, where
the kinematic peculiarity is seen only in the \ion{H}{1} disk; and

\noindent (5) Reference for the data presentation and discussion.

For each object, an {\it R\/}--band 
image was obtained with the Kitt Peak\footnote{The Kitt Peak
National Observatory is a facility of the National Optical Astronomy
Observatories, which is operated by the Association of Universities
for Research in Astronomy, Inc. (AURA) under a cooperative agreement
with the National Science Foundation.} 0.9~meter telescope and a
21~cm \ion{H}{1} line synthesis map was made with the
VLA in either the C or D configuration. 
For details of the datasets and their acquisition and reduction
procedures, see Papers~I---III and {\S}3.3.2 of Jore (1997). 
Here we discuss only those aspects that are specific to the current
work.

\section{Data Analysis}

\subsection{Obtaining Derived Rotation Curves}

The moment maps representing the \ion{H}{1} velocity fields obtained
from Papers~I---III were analyzed in the GIPSY environment (van der
Hulst {\it et al.\/} 1992\markcite{gipsy}, {\it GIPSY\/}
2000\markcite{gip2000}). Rotation curves were derived by iteratively
applying the standard GIPSY routine ROTCUR (Begeman
1989)\markcite{rotcur}. This routine divides the galaxy into a number
of rings, and fits a function
\begin{equation}
V(x,y) = V_{sys} + V_{rot} \cos\theta \sin i
\end{equation}
where
\begin{equation}
\cos\theta = \frac{-\left(x-x_0\right) \sin(\Gamma) + \left(y-y_0\right) \cos(\Gamma)}{r}
\end{equation}
to the velocity field in each ring, allowing the systematic velocity
$V_{sys}$, the rotation velocity $V_{rot}$, inclination $i$, position
angle $\Gamma$, and kinematic center $\left(x_0,y_0\right)$ to
vary. The implicit assumption is that the gas falls in nearly flat, 
circular orbits. The normal morphology of the galaxies studied here 
suggests that this is likely the case.

Because the \ion{H}{1} in NGC~3626 is made up of independent inner and
outer rings (see Paper~II), these rings were fit independently of one
another, and later combined into a single model.

The tilted--ring model fits were obtained iteratively in the following
way. The velocity fields of each galaxy were fit, using a $\cos
\theta$ weighting function, and allowing all parameters to vary. The
purpose of this first fit was to find the kinematic center or centers
of the galaxy. In some cases, the kinematic center differed by more
than a beam width between the inner and outer regions of the disk. In
these cases, two independent centers, $c_1$ and $c_2$ were fit to the
inner and outer regions, respectively. The derived kinematic centers
are presented in Table~\ref{Kornreich:tab2}.

\begin{deluxetable}{llllll}
\tablenum{2}
\tablewidth{6 in}
\tablecaption{Derived Kinematic Centers}
\tablehead{
\colhead{Galaxy}&
\multicolumn{2}{c}{$c_1$ (1950)}&
\multicolumn{2}{c}{$c_2$ (1950)}&
\colhead{$R_b$}
\\[.2ex]
\colhead{}&
\colhead{$\alpha$}&
\colhead{$\delta$}&
\colhead{$\alpha$}&
\colhead{$\delta$}&
\colhead{(arcsec)}
}
\startdata

NGC~1169 &	030011.1 & +461129.3 &	030011.9 & +461140.0 & 	240 \\
NGC~3623 &	111619.2 & +132157.9 &    \nodata& \nodata   & \nodata \\
NGC~3626 &	111726.2 & +183746.4 & 	111725.9 & +183752.6  & 52 \\
NGC~3900 &	114633.8 & +271759.9 &	\nodata& \nodata   & \nodata \\
NGC~4138 &	120658.1 & +435748.9 &	120659.4 & +435730.9 & 140 \\
NGC~4772 &	125056.2 & +022614.9 &    \nodata& \nodata   & \nodata \\
NGC~4866 &	125658.3 & +142623.9 &    \nodata& \nodata   & \nodata \\
NGC~5377 &	135417.8 & +472845.7 &    \nodata& \nodata   & \nodata \\
NGC~5448 &	140055.9 & +492443.3 &    \nodata& \nodata   & \nodata \\

\enddata
\label{Kornreich:tab2}
\end{deluxetable}

The columns represent the galaxy designation, the right ascension and
declination of the derived kinematic center $c_1$ (or kinematic center
of the inner disk, where two centers were fit), the kinematic center
of the outer disk $c_2$, where two centers were fit, and the boundary
radius $R_{b}$ between the inner and outer fits. In all cases, the
derived (inner) center $c_1$ agrees with the kinematic centers derived
in Papers~I---III using a slightly different approach.

\begin{figure}
\plotfiddle{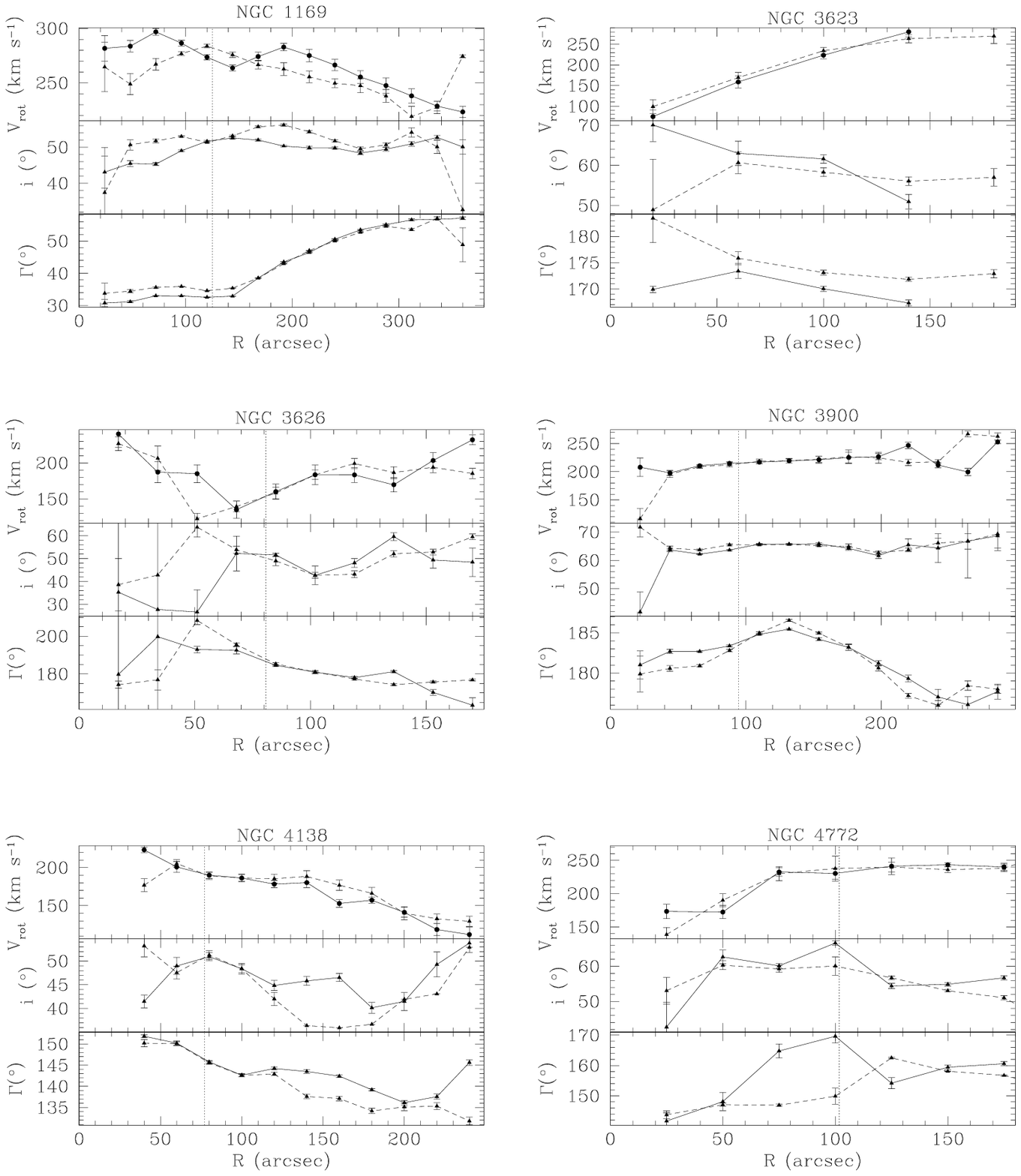}{4.0in}{0}{60}{60}{-180}{-90}
\plotfiddle{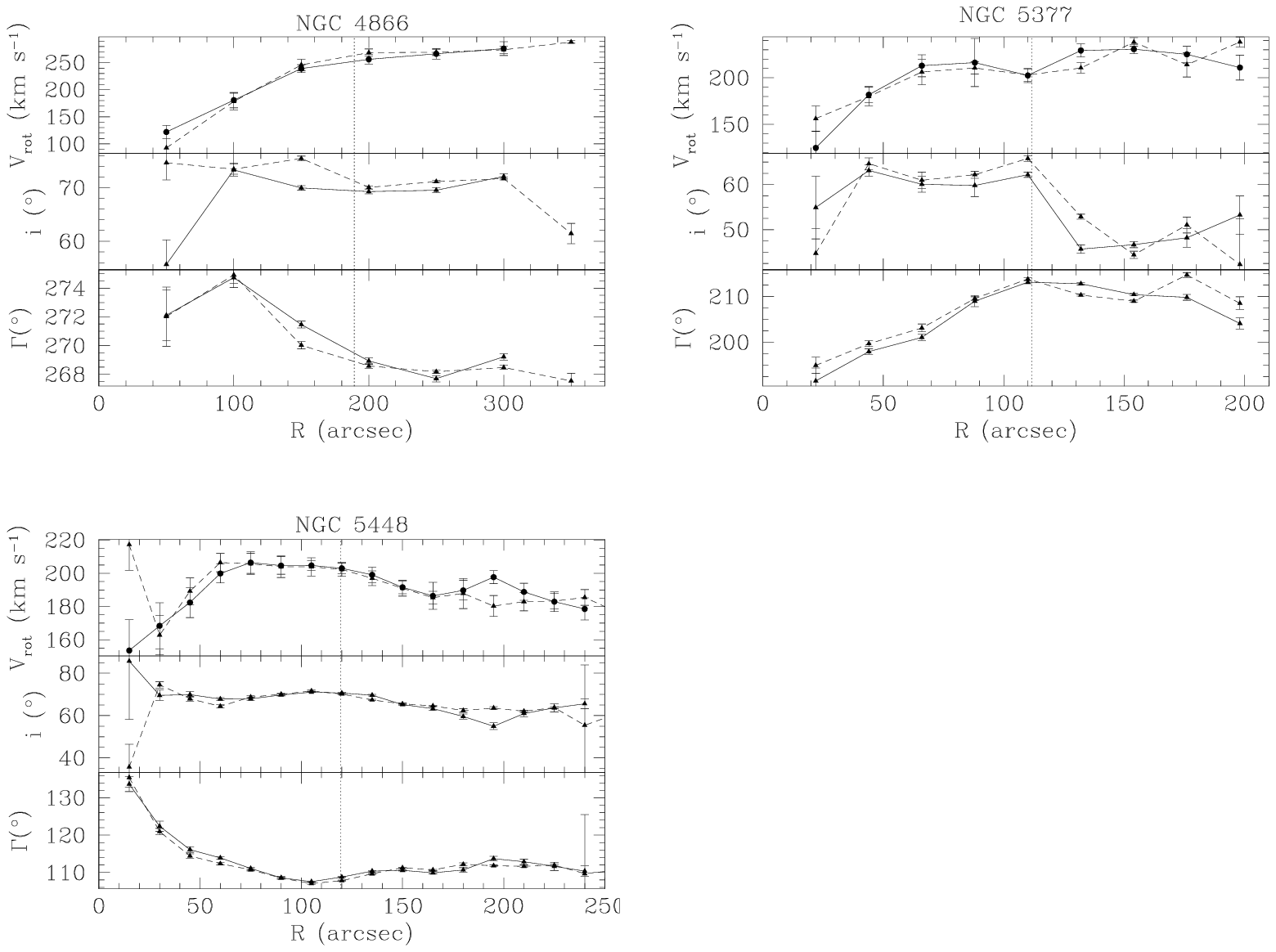}{4.0in}{0}{60}{60}{-180}{-60}
\vskip -3cm
\caption[]{
Rotation curves and position
angle plots derived from the tilted--ring fitting for the nine
galaxies in the Sa sample. The $x$ axis of each frame is the radius of
the galaxy in arcseconds, while the $y$ axes of each frame, are, top
to bottom, the derived rotation $V_{\it rot}$, the derived kinematic
inclination $i$ in degrees, and the derived kinematic position angle
$\Gamma$, also in degrees. Solid lines represent the fit for the
approaching side; dashed lines that of the receding side. The vertical
dotted line in each frame represents $R_{25}$.
\label{Kornreich:fig1}}
\end{figure}

Following the fit of the kinematic centers, the approaching and
receding sides of the galaxy were fit separately, with the kinematic
centers fixed. In cases where two centers were derived, inner ($r \leq
R_b$) and outer ($R>R_b$) regions were fit independently. The fit
obtained for both sides used as the initial guess in the
iteration. The results of these fits then represented the rotation
velocity $V_{rot}$, the inclination $i$, and the position angle
$\Gamma$ as a function of radius. These fits are presented in
Figure~\ref{Kornreich:fig1}.  For each galaxy frame, the radius is given 
on the $x$ axis in arcseconds, while $V_{rot}$, $i$, and $\Gamma$ are
given on the $y$ axis in km~s$^{-1}$, degrees, and degrees,
respectively. Circles and solid lines trace the approaching side,
while triangles and dotted lines trace the receding side.

\subsection{Construction of Tip--LON Diagrams}

Briggs (1990)\markcite{briggs} has employed a graphical representation
of position angle versus inclination as functions of radius to
illustrate the degree and character of warping in \ion{H}{1} disks. Because
the tilted--ring fits in this work were accomplished with
independent $V_{rot}$ and $i$, it was possible to represent the
changes in the dynamic structure of the galaxies via these
position--angle/inclination diagrams. These ``tip--LON'' diagrams
are useful for studying the structure of galactic warping and
kinematic bending of the disk. The plots represent the kinematic
orientation (or ``tip angle'') of the rings, as a function of radius,
referred back to the average plane of the galaxy. 

All of the information of the position angle $\Gamma$ and inclination
$i$ at radius $R$ is contained within the vector normal to the plane
of the fitted ring. If we define a hemisphere with the galaxy at the
center of the circular base, and where the base represents the plane
of the sky as seen from Earth, such that the telescope lies directly
above the apex of the hemisphere, then the set of points of
intersection of the hemisphere with the normal vectors will also
contain all of the information contained within $\Gamma$ and $i$.  The
resulting points are located at a ``latitude'' equal to $90^\circ-i$,
and ``longitude'' $\Gamma+90^\circ$ (See Figure~1 of Briggs 1990).
The azimuthal coordinate corresponds to the direction 90$^{\circ}$
from the line of nodes (LON) of the ring in the reference plane, hence
the name ``tip--LON'' diagram adopted by Briggs.

Once the points on the hemispherical surface are determined, the
points are then projected onto a plane, tangent to the sphere at the
point representing the first ring, or at some other convenient nearby
point. The final result is a spherical projection mapping the
kinematic orientation for all radii.  For a more detailed description
of the mapping procedure, see Briggs (1990).

\begin{figure}[h]
\plotfiddle{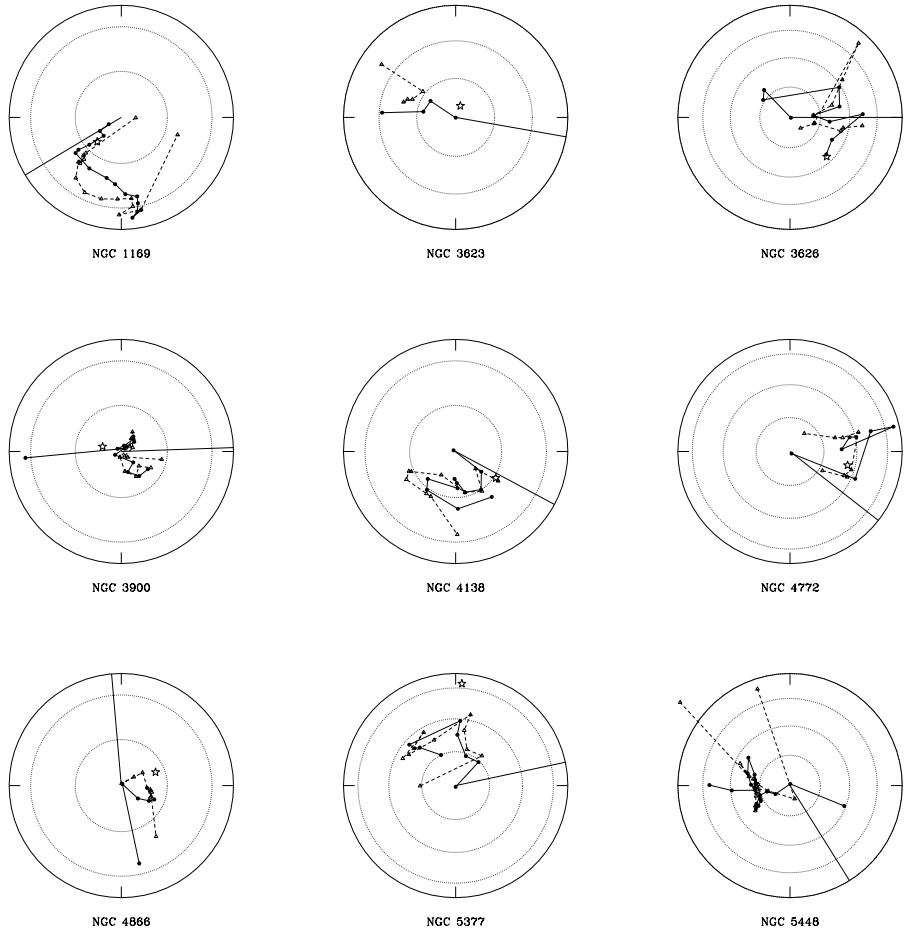}{3.0in}{0}{70}{70}{-210}{-240}
\vskip 4.8cm
\caption[]{Tip--LON diagrams for the nine galaxies in the sample. Both the
approaching (circles, solid lines) and receding (triangles, dashed
lines) are presented. The open star in each diagram represents the
morphological tip angle derived from the {\it R\/}--band optical
isophote at $D_{25}$. The solid line extending from the center of the
diagram to the outer circle indicates the direction away from the
$i=0^{\circ}$ pole, indicating the direction of increasing
inclinations. Successive dotted concentric circles are of 10$^\circ$
deviation each.\label{Kornreich:fig2}}
\end{figure}

The tip--LON diagrams for the current inclined Sa galaxy sample are
presented in Figure~\ref{Kornreich:fig2}. In these diagrams, as in the
rotation curves, circles and solid lines represent the approaching
side of the galaxy; triangles and dashed lines the receding side. Also
included in each diagram is a single point, represented as an open
star, indicating the observed morphological tip angle of the optical
galaxy as determined by isophotal ellipse fitting of the isophote at
$R_{25}$ in the {\it R\/}--band images from Papers I---III as
performed using the GALPHOT\footnote{The GALPHOT surface photometry
package is a collection of IRAF/STSDAS scripts originally developed by
Wolfram Freudling and John Salzer; the Cornell version has been
further modified and is maintained by M. P. H.} package of
IRAF\footnote{IRAF is distributed by the National Optical Astronomy
Observatory.}/STSDAS\footnote{STSDAS (Space Telescope Science Data
Analysis System) is distributed by the Space Telescope Science
Institute which is operated by AURA under contract to the National
Aeronautics and Space Administration.}. The solid line radiating from
the center of the plot indicates the direction away from the line of
sight, such that deviations roughly parallel to this line represent
deviations in $i$. Deviations perpendicular to this line represent
mostly deviations in $\Gamma$. The large, concentric, dotted circles
each represent $10^\circ$ deviation (or ``tip'') from the tangent
point, located at the center of the diagram. Note that each diagram is
cut off at a different maximum deviation. For each diagram, the
tangent point is chosen to be the first fitted ring point on the
approaching side, except where this point is of large uncertainty or
is isolated from the rest of the plotted points. In those cases, the
point of tangency is chosen to be the second plotted point on the
approaching side. A disk rotating in a flat plane, free of warps and
streaming motions, should have all its points concentrated in the
center of the diagram.

\subsection{Analysis of Nonaxisymmetries}

In KHL and KHLvZ, we discussed methods for quantifying the
morphological and dynamical asymmetry of a galaxy. Morphological
asymmetry can be measured in either the optical or \ion{H}{1} regimes
using the ``Method of Sectors'' described in KHL. First, some
appropriate center of the galaxy is determined; this center can be the
optical center of light, the dynamical center, or some other
appropriate choice. Radials are then drawn outward from the center,
dividing the galaxy into a number $n$ (usually 8) of sectors. Inner and
outer boundary radii are then selected inside of which flux
integration will be performed. Generally, the inner radius is selected 
to leave out contributions from the bulge or bar components, while the 
outer radius is selected to include as much flux from the galaxy as
possible while minimizing noise. The flux is then integrated inside
each sector. The maximum flux difference is then normalized to the
total measured flux to produce the asymmetry measure
\begin{equation}
A_f = \frac{f^{\it max} - f^{\it min}}{\sum_n f_n},
\end{equation}
where $f^{\it max}$ and $f^{\it min}$ are the maximum and minimum flux
detected in a sector, and $f_n$ is the flux in the nth sector.  For
galaxies observed at inclinations other than face--on, the sector
pattern can be projected to the appropriate angle.

Quantifying dynamical asymmetry was performed in KHLvZ by examining
the rotation curves of target galaxies. In a perfectly axisymmetric
disk, approaching and receding rotation curves should be identical, and 
the kinematic position angle should be constant as a function of
radius and between the approaching and receding sides. The measure of
rotation curve symmetry is performed by superposing the approaching
and receding sides of the rotation curve and integrating the
differences between them, normalized by half of the velocity width. In 
KHLvZ, measurements were taken using the observed rotation curves:
\begin{equation}
S_2^{\it obs} \equiv 10^2 \times \frac{\int \left|\left|V_{rot}(r)\sin i\right| -
\left|V_{rot}(-r)\sin i\right|\right|\,dr}{\int \frac12\left[
\left|V_{rot}(r)\sin i\right| + \left|V_{rot}(-r)\sin i\right|\right]\,dr}.
\label{S2EQ}
\end{equation}

To quantify the deviations from constant position angle, two
measurements were taken. First, the maximum difference in position
angle as a function of radius
\begin{equation}
\left.\Delta\Gamma(r)\right|_> = {\rm max}\,|\Gamma(r_n) - 
\Gamma(r_m)|,
\end{equation}
where $n$ and $m$ are indices of rings in the tilted--ring fit on the 
same side (approaching or receding) of the fit, and second, the
average of position angle differences between sides:
\begin{equation}
\Delta\Gamma(\theta) = \frac1n \sum_n \left| \Gamma(-r_n) -
\Gamma(r_n) \right|.
\end{equation}
Note that the latter two parameters are introduced simply as
$\Delta\Gamma(r)$ and $\Delta\Gamma(\theta)$ in KHLvZ.

Finally, Haynes \etal (1998)\markcite{HHMRvZ} describe the measurement
of asymmetry in \ion{H}{1} line profiles. The quantity defined there
is:
\begin{equation}
A_{l/h} \equiv \frac{\int_{V_l}^{V_{\rm med}} S\,dV}{\int_{V_{\rm
med}}^{V_h} S\,dV}
\end{equation}
where $V_l$ and $V_h$ are the low and high velocity limits,
respectively, of signal detection above $3\sigma$, and $V_{\rm med}$
is the median between them. The characteristic lopsidedness of the
line profile is given simply by
\begin{equation}
A_n \equiv \cases{A_{l/h},&if $A_{l/h}>1$\cr
		  1/A_{l/h},&otherwise.\cr}
\end{equation}

Following the discussion in KHLvZ, and using
the fitted rotation curves, the dynamical parameters of the observed
rotation curve asymmetry $S_2^{\it obs}$
(equation~\ref{S2EQ}), the derived rotation curve asymmetry
\begin{equation}
S_2^{\it drv} \equiv 10^2 \times \frac{\int \left|\left|V_{rot}(r)\right| -
\left|V_{rot}(-r)\right|\right|\,dr}{\int \frac12\left[
\left|V_{rot}(r)\right| + \left|V_{rot}(-r)\right|\right]\,dr} ,
\label{Kornreich:S2EQ}
\end{equation}
the maximum change in
position angle with radius $\left.\Delta\Gamma(r)\right|_>$, and the
average difference in position angle between approaching and receding
sides at constant radius $\left<\Delta\Gamma(\theta)\right>$ were
calculated.  Here, we expand this analysis to calculate
the average difference in position angle with radius on the same side,
$\left<\Delta\Gamma(r)\right>$, and the maximum change in position
angle at constant radius between sides,
$\left.\Delta\Gamma(\theta)\right|_>$. 

The zeroth moment maps representing the \ion{H}{1} column density
obtained from Papers~I---III were also analyzed in the GIPSY
environment following the discussion of morphological asymmetry
quantification described in KHLvZ, using a total of eight segments
each. Because all of the galaxies in this sample are inclined, the
segment pattern had to be projected to the position angle and
inclination of the target galaxies, averaged over all radii. The
segments were numbered 1 through 8, with segment number 1 being
located just north of the western semi--major axis, and with the
remaining segments numbered consecutively in a counter--clockwise
fashion. No corrections were made for internal extinction.

The {\it R\/}--band images were analyzed for morphological asymmetry using
the ``Method of Sectors'' described in KHL. Foreground stars and other interfering
objects were removed from the optical frame by replacing them with a
second-order polynomial interpolation of the surrounding background
before the segment analysis.

Finally, the synthesized \ion{H}{1} line profiles  obtained
from Papers~I---III were analyzed for asymmetry following the
discussion of Haynes \etal (1998)\markcite{HHMRvZ}.

The results of the asymmetry quantifiers are presented in
Table~\ref{Kornreich:tab3}. The columns in this table represent:

\noindent (1) The NGC designation of the galaxy;

\noindent (2) The morphological asymmetry $A_{f,R}$ obtained from the 
segmented flux comparison in the {\it R\/}--band. Following the value
of $A_{f,R}$, the maximum and minimum flux segments are identified by
number in parentheses;

\noindent (3) The morphological asymmetry $A_{f,HI}$ obtained from the 
segmented flux comparison in \ion{H}{1}. Following the value of $A_{f,HI}$, the
maximum and minimum flux segments are identified by number in
parentheses;

\noindent (4) The observed rotation curve asymmetry $S_2^{\it obs}$

\noindent (5) The derived rotation curve asymmetry $S_2^{\it drv}$;

\noindent (6) The mean difference in position angle with radius on
the same side, $\left<\Delta\Gamma(r)\right>$;

\noindent (7) The mean difference in position angle on opposite sides
at the same radius, $\left<\Delta\Gamma(r)\right>$;

\noindent (8) The greatest difference in position angle with radius on
the same side, $\left.\Delta\Gamma(r)\right|_>$;

\noindent (9) The greatest difference in position angle on opposite
sides at the same radius, $\left.\Delta\Gamma(r)\right|_>$; and

\noindent (10) The asymmetry in the synthesized \ion{H}{1} line profiles, $A_n$.

\begin{deluxetable}{lrrrrrrrrr}
\scriptsize
\tablenum{3}
\tablecaption{Asymmetry Measures of the Sa Sample}
\tablehead{
\colhead{Galaxy}&
\colhead{$A_{f,R}$}&
\colhead{$A_{f,HI}$}&
\colhead{$S_2^{\it obs}$} &
\colhead{$S_2^{\it drv}$} &
\colhead{$\left<\Delta\Gamma(r)\right>$} &
\colhead{$\left<\Delta\Gamma(\theta)\right>$} &
\colhead{$\left.\Delta\Gamma(r)\right|_>$} &
\colhead{$\left.\Delta\Gamma(\theta)\right|_>$} &
\colhead{$A_n$}
\\[.2ex]
\colhead{} &
\colhead{} &
\colhead{} &
\colhead{} &
\colhead{} &
\colhead{($^\circ$)} &
\colhead{($^\circ$)} &
\colhead{($^\circ$)} &
\colhead{($^\circ$)} &
\colhead{} 
}
\startdata

NGC~1169 & 0.0248 (4,8) &0.0534 (6,3) &1.20    & 2.87    & 1.8 $\pm$ 0.12  & 2.25 $\pm$ 1.7  & 26.3 $\pm$ 1.3  & 9  $\pm$ 0.3  & 1.01\cr
NGC~3623 & 0.0441 (3,6) &0.1110 (3,1) &2.42    & 7.32    & $-$1.9 $\pm$ 1.0  & 13.5 $\pm$ 4.7  & 11.6 $\pm$ 4.6  & 5.9 $\pm$ 2.0   & 1.11\cr 
NGC~3626 & 0.0208 (2,4) &0.1086 (5,3) &8.77    & 9.88    & $-$1.1 $\pm$ 5.4  & 23 $\pm$  8     & 36   $\pm$ 10  & 7.4 $\pm$ 12    & 1.06\cr
NGC~3900 & 0.0123 (5,2) &0.0526 (8,7) &5.33    & 6.36    & $-$0.2 $\pm$ 0.3  & 2.4 $\pm$ 1.2   & 10.5 $\pm$ 0.6  & 1.1 $\pm$ 0.7   & 1.15\cr
NGC~4138 & 0.0200 (1,3) &0.0448 (6,3) &3.91    & 5.92    & $-$1.3 $\pm$ 0.4  &18.4 $\pm$ 1.1   & 18.7 $\pm$ 1.2 & 2.2 $\pm$ 1.0   & \nodata\cr
NGC~4772 & 0.0152 (7,3) &0.0809 (5,2) &4.26    & 4.11    & 2.4 $\pm$ 1.0      & 19.7 $\pm$ 3.5  & 27.9 $\pm$ 3.5 & 7.8 $\pm$ 2.2   & 1.16\cr
NGC~4866 & 0.0120 (7,4) &0.1332 (8,6) &3.76    & 3.50    & $-$0.89 $\pm$ 0.4   & 1.42 $\pm$ 0.35 & 7.0  $\pm$ 0.7& 0.5 $\pm$ 0.8   & 1.15\cr
NGC~5377 & 0.0496 (8,6) &0.1584 (8,2) &1.92    & 4.91    & 1.65 $\pm$ 0.5    & 4.9  $\pm$ 0.8  & 21.6 $\pm$ 2.8 & 2.4 $\pm$ 1.1   & 1.10\cr
NGC~5448 & 0.0262 (1,4) &0.0985 (7,6) &1.50    & 3.00    & $-$1.3 $\pm$ 0.8  & 1.9  $\pm$ 0.7  & 28.4 $\pm$ 2.5 & 1.0 $\pm$ 1.7   & 1.06\cr

\enddata
\label{Kornreich:tab3}
\end{deluxetable}

\section{Discussion}

\subsection{Asymmetry Characteristics of the Sa Sample}

The Jore (1997) sample of Sa galaxies was selected based on normal
optical morphology; therefore, it is not surprising to find that the
measurements of optical asymmetry for these galaxies are small when
compared to similar measurements of lopsided galaxies in
KHLvZ. Only two galaxies in this sample exhibit
significant asymmetry ($A_{f,R}>0.03$) as measured by this method: NGC~3623 and
NGC~5377. The asymmetry measurement for NGC~3623 reflects the prominent
dust lanes observed in the eastern portions of the galaxy, while that
of NGC~5377 reflects the presence of a low--surface--density outer
disk at an apparent position angle different from that of the inner disk.

This apparently quiescent story changes when the \ion{H}{1} disks
are examined. Large values of the \ion{H}{1} morphological asymmetry
$A_{f,HI}$ and the various $\Delta\Gamma$ parameters are observed
throughout the sample. Most values of $S_2$ detected in this sample do 
not rise to the amplitude of those detected in KHLvZ, 
but this may reflect two significant effects: first, the difficulty of 
obtaining adequate rotation curves for the face--on galaxies of
KHLvZ, and second, that large values of $S_2$
in face--on objects primarily reflect non--planar motions, while in
inclined objects they reflect non--circular motions. Therefore, these
results indicate that non--circular motions in the Sa sample are of
lesser magnitude than the non--planar motions in the face--on sample.

Nevertheless, the asymmetry detected in the rotation curves of these
kinematically disturbed galaxies is still of order 4--10\%,
significantly larger than the 1--2\% asymmetry in the $n=2$ spiral
mode measured by SFdZ for the
kinematically normal galaxies NGC~2403 and NGC~3198.

The dichotomy between the small asymmetries measured in the optical
and the large asymmetries measured in the \ion{H}{1} is not surprising; it
only confirms the reports of kinematic decoupling discussed in
Papers~I---III. Thus these findings suggest the hypothesis that the
stellar component, evident in the optical images,
recovers from the disruptive event which caused the
kinematic decoupling more quickly than the hydrodynamic components, and that
galaxies with disturbed kinematics can still appear morphologically
normal.

\subsection{Evaluation of Briggs' Rules for Warps in the Sa
Sample}

Briggs (1990) has studied the characteristics of warps in a sample of 12
galaxies whose kinematics were determined from \ion{H}{1} synthesis
maps. From his sample, he inferred four rules of behavior for warps in
galactic disks: first, that disks are planar within $R_{25}$, but
warping in inclination becomes evident at the Holmberg radius $R_{\it Ho}
\equiv R_{26.5}$. Second, that warps change character at a transition
radius near $R_{\it Ho}$. Third, that beyond $R_{\it Ho}$, the position angle
advances in the direction of galactic rotation. Finally, fourth, that
gas far beyond $R_{\it Ho}$ defines a line of nodes differently oriented
from that of the inner disk. Here, we comment on the adherence of our
sample to these rules of warps.

The characteristic signature in the tip--LON diagram of a galaxy which
follows the above rules is a clustering of points representing radii
$r<R_{25}$ in the center of the diagram, followed by the outer points
spiraling outward to larger tip angles in the direction of galactic
rotation. Here, we have typically not been able to determine the direction of
rotation, but can see that NGC~1169 clearly shows the signature in the
other particulars. The galaxy is roughly planar, though slightly
increasing in inclination, to $R_{25} = 125^{\prime\prime}$. Beyond
about $r=140^{\prime\prime}$, however, the position angle rapidly and
smoothly increases by 25$^\circ$.

This signature is not as clearly seen in any of the remaining
galaxies. NGC~3623 exhibits a gradual inclination decrease over the
entire disk, but shows little position angle change. NGC~3626 exhibits
steadily changing position angle but steady inclination throughout the
truncated disk, well within $R_{25}$. NGC~3900 exhibits no more than
10$^\circ$ of tip at any radius, except for the first point on the
approaching side, which has large errors. NGC~4138 exhibits a
spiralling pattern, but one which changes direction abruptly at $\sim
2R_{25}$. NGC~4772 exhibits a spiraling pattern, but there is no
apparent transition radius. NGC~4866 shows no warping. NGC~5377
exhibits two regimes: one within $R_{25} = 111^{\prime\prime}$ in
which the position angle smoothly increases, and another beyond
$R_{25}$ in which there is no change in $i$ or $\Gamma$, but in which
these values are different than those of the inner disk. It is also
interesting to note that the optical morphological tip angle in
NGC~5377, as represented by the starred point in
Figure~\ref{Kornreich:fig2}, is the only one which does not correspond
well with the kinematic tip angles in \ion{H}{1}. This discrepancy is
due to the presence of separate inner and outer stellar disks, which
exhibit very different inclinations and position angles. The
\ion{H}{1} seems to exhibit tip angles similar to the outer disk,
while $R_{25}$ is well within this inner disk. Finally, NGC~5448
exhibits some 20$^\circ$ of position angle change within radii $r \leq
.6R_{25}$, but is not warped beyond $R_{25}$.

The only galaxy in this sample which clearly follows Briggs' rules,
therefore, is the barred SBa galaxy NGC~1169. It is important to note,
however, that NGC~1169 is one of only two galaxies in the sample which
does not show evidence of peculiar kinematics. The other, NGC~5448, is
not warped outside of the inner disk; therefore, we should not expect
the rules to apply. Thus, the two kinematically normal galaxies in the
sample are consistent with Briggs' rules for warps. The remaining,
kinematically decoupled galaxies, where they exhibit warping, are
inconsistent with these rules as a group. While most exhibit warps of some
kind, the warps are qualitatively different from the warps observed
in kinematically normal galaxies.

\subsection{Searching For Trends in Asymmetry}

It is important to note for the purposes of this section that the
galaxies selected for these observations were mostly gas rich, with \ion{H}{1}
masses ranging from $10^9 \lesssim M_{HI} \lesssim 6\times
10^9M_\odot$, compared to the range of $5\times10^8
\lesssim M_{HI} \lesssim 6\times 10^9M_\odot$ for Sa galaxies in the
Local Supercluster as tabulated by Roberts \& Haynes
(1994)\markcite{rh94}. They also represent a very narrow and red
range of colors, with nearly all \bv values near 0.82, compared to
ranges of 0.6 $\lesssim$ \bv $\lesssim$ 0.8 given in Roberts \& Haynes
as representative of early--type disks. This subset is also 
small, consisting of only nine objects. Therefore, this sample is not
representative of Sa galaxies as a class, and any relationship
observed among physical parameters here or any extrapolation to Sa
galaxies as a whole should be approached with the utmost caution.

\begin{figure}[h]
\plotfiddle{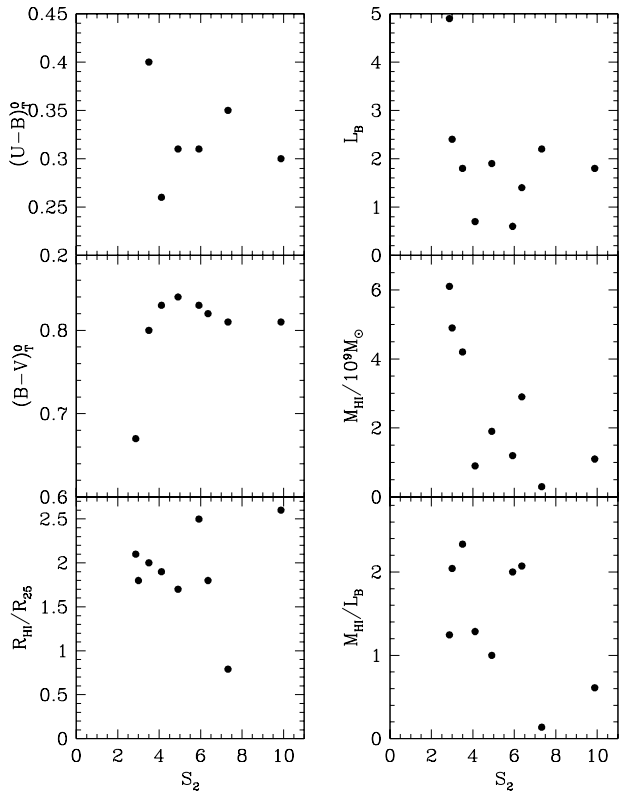}{3.0in}{0}{70}{70}{-220}{-210}
\vskip 2.8cm
\caption[]{Plots of intrinsic physical properties of the Sa 
sample, as a function of the derived rotation curve asymmetry
$S_2$. From top to bottom, left to right: the \ub\ color index, corrected for
external and galactic extinction and for redshift from the RC3; the
\bv\ color index, corrected for external and galactic extinction and for
redshift from the RC3; the radius at $N_{HI}=1M_\odot {\rm pc}^{-2}$,
$R_{HI}$, in terms of $R_{25}$; total blue luminosity in terms of
$10^{10}L_\odot$ from Papers~I---III; the total \ion{H}{1} mass in terms of
$10^9M_\odot$ from Papers~I---III; and the total \ion{H}{1} mass to blue
luminosity ratio in terms of $M_\odot/L_\odot$.\label{Kornreich:fig3}}
\end{figure}

In order to examine the relationships between the kinematic
asymmetries observed in the sample and the physical properties of the
galaxies, the derived rotation curve asymmetry measure $S_2^{\it drv}$
presented in Table~\ref{Kornreich:tab3} was compared to the global optical
and \ion{H}{1} properties of the sample galaxies presented by
Papers~I---III. These physical properties were: the \ub\ color index,
the \bv\ color index, the relative size of the \ion{H}{1} disk at a
level of 1~$M_\odot$~pc$^{-2}$ to the optical disk $R_{HI}/R_{25}$,
the total blue luminosity $L_B/10^{10}M_\odot$, the total \ion{H}{1}
mass $M_{HI}/10^9M_\odot$, and the \ion{H}{1} mass to blue luminosity
ratio $M_{HI}/L_B$. Plots of pairs of these parameters are presented
in Figure~\ref{Kornreich:fig3}.

\vskip 10pt

It appears that galaxies in this sample with smaller \ion{H}{1}
masses tend to have greater asymmetry measures. That there is no
similar trend with $R_{HI}/R_{25}$ indicates that the issue is
not a small number of points measured along the rotation curve, whose
errors would tend to inflate the measurement of $S_2$. It may
indicate that smaller galaxies have smaller overall \ion{H}{1} masses, making them
more susceptible to perturbation. However, the small range of masses
probed makes the significance of this tendency uncertain.

\begin{figure}[h]
\plotfiddle{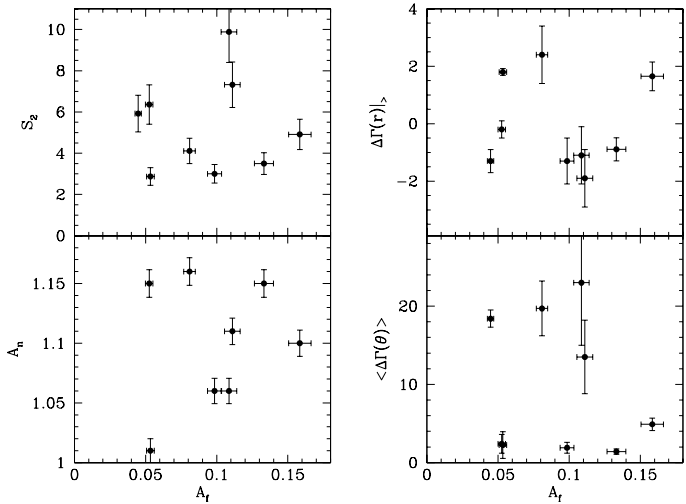}{3.0in}{-90}{70}{70}{-250}{340}
\vskip -0.7cm
\caption[]{Illustration 
of the relationships between the morphological asymmetry parameter
$A_f$ and dynamical asymmetry parameters $S_2$, the rotation curve
asymmetry, $A_n$, the line profile asymmetry,
$\left.\Delta\Gamma(r)\right|_>$, the greatest position angle asymmetry in
radius, and $\left<\Delta\Gamma(\theta)\right>$, the average position
angle asymmetry between approaching and receding sides.\label{Kornreich:fig4}}
\end{figure}

The various quantifiers of morphological and dynamical asymmetry
derived in Table~\ref{Kornreich:tab3} were also compared to each other,
again to search for trends among the various
parameters. Figure~\ref{Kornreich:fig4}  illustrates the relationships
between the morphological asymmetry parameter $A_f$ in the \ion{H}{1} to
dynamical parameters. No significant trends are detected between
morphological and dynamical parameters. The very weak correlation
observed in KHLvZ between $A_f$ and
$\left<\Delta\Gamma(\theta)\right>$ is not recovered. 

\vskip 10pt

The physical meaning of this apparent non--correlation is not entirely
clear, however. The Sa galaxy sample (Jore 1997) was selected in part based
on moderate inclination, which allows the dynamical parameters to be
well--constrained, but introduces uncertainty in the determination of
the morphological parameters. Furthermore, Sa galaxies are often
dusty and show tightly wound spiral structure with patchy star
formation regions. Conversely, the face--on aspect of the
sample of KHL and KHLvZ constrains the morphological
quantifiers, but introduces uncertainty in the dynamical ones.

\begin{figure}[h]
\plotfiddle{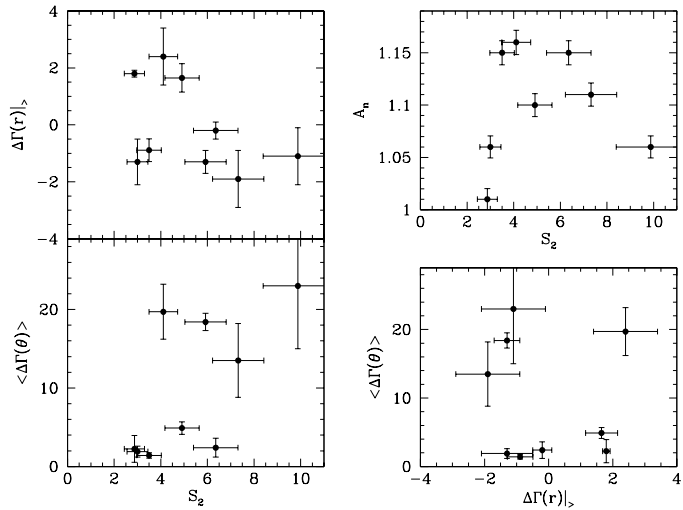}{3.0in}{-90}{70}{70}{-270}{350}
\vskip -1.2cm
\caption[]{Illustration 
of the relationships among the various dynamical asymmetry parameters,
including the derived rotation curve asymmetry $S_2^{\it drv}$, $A_n$,
the line profile asymmetry, $\left.\Delta\Gamma(r)\right|_>$, the
greatest position angle asymmetry in radius, and
$\left<\Delta\Gamma(\theta)\right>$, the average position angle
asymmetry between approaching and receding sides.\label{Kornreich:fig5}}
\end{figure}

Figure~\ref{Kornreich:fig5} illustrates the relationships among the
dynamical asymmetry parameters. The strongest observed trend is
between the derived rotation curve asymmetry $S_2^{\it drv}$ and the
average position angle change between sides,
$\left<\Delta\Gamma(\theta)\right>$, although even this is only
marginally detected. This trend could reflect imprecise centering,
which would inflate both values together, but such improper centering
would also inflate measures of $A_f$ and $\Delta\Gamma(r)$. As no
correlation is detected between these values and either $S_2$ or
$\left<\Delta\Gamma(\theta)\right>$, improper centering can be
eliminated as a source of this effect.

This relationship therefore is likely a real physical effect. The
simplest explanation of the correlation is that localized streaming
motions, which affect the rotation curve and position angle on one
side of a galaxy, are detected both in $S_2$ and
$\left<\Delta\Gamma(\theta)\right>$. More exotic explanations,
however, such as mass infall or minor mergers, cannot be ruled out.

\subsection{Comparing $S_2$ with Global Kinematic Asymmetry}

Swaters \etal (1999)\markcite{S3vA} have found that kinematic lopsidedness
as measured by Fourier decomposition was intimately connected with
qualitatively asymmetric rotation curves. To test this hypothesis, we
apply the method of SFdZ to our sample of nine galaxies to find the zeroth
through third harmonic components of each galaxy's velocity field as a
function of radius. In this analysis, only rings with filling factors
in excess of 0.5 were considered, to reduce the effects of noise in
the outermost rings.

SFdZ define the average amplitude in the $i$th harmonic component as
\begin{equation}
\sum_j \left[\hat c_i(j)^2 + \hat s_i(j)^2 \right]^{1/2}/N,
\end{equation}
where each of $N$ sampled radii is labeled by $j$, and the $\hat c_i$ and
$\hat s_i$ are defined by:
\begin{equation}
v_{\rm los} = v_{\rm sys} + \sum_n \hat c_n\cos n \hat\psi +
\hat s_n\sin n \hat\psi,
\end{equation}
where $\hat \psi$ is the azimuthal angle from the major axis in the
plane of the galaxy. The exception is the $i=1$ harmonic, in
which $\hat c_1(r)$ is $V_{rot}$, and the first harmonic component is
taken to be the average of the $\hat s_1$ terms. The error analysis
conducted here follows the discussion of SFdZ, but it is important to
note that the error bars presented in the following figures represent
only the formal numerical error. Systemic errors due to random
velocities in gas clumps may also be of similar order.

\begin{figure}[h]
\plotfiddle{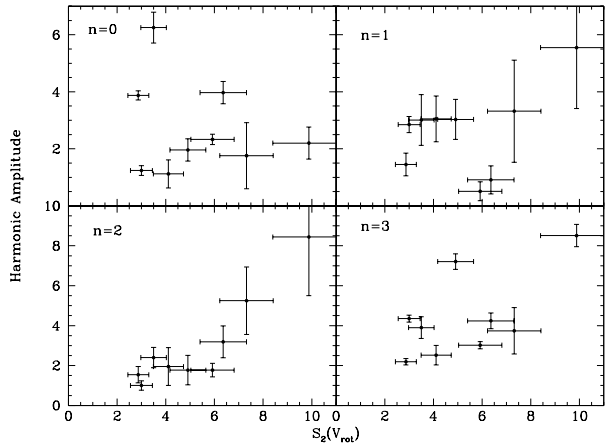}{3.0in}{-90}{70}{70}{-270}{350}
\vskip -1.2cm
\caption[]{Illustration 
of the relationships between the derived rotation curve asymmetry
parameter $S_2^{\it drv}$ and the $n={0,1,2,3}$ dynamical harmonic components as
measured by the method of SFdZ.\label{Kornreich:fig6}}
\end{figure}

These results were then plotted against our measure of rotation curve
asymmetry $S_2^{\it drv}$. The results are presented in Figure
\ref{Kornreich:fig6},  in which each frame traces a harmonic component 
as a function of $S_2$ in the sample. In agreement with Swaters {\it
et al.\/}  (1999), we find that the rotation curve asymmetry measure $S_2$ is
correlated with the $n=2$ Fourier mode. Trends of increasing $n=1$ and
$n=3$ are also observed for increasing $S_2$. In addition, $S_2$ values for
the rotation curves of the kinematically asymmetric galaxies DDO~9 and
NGC~4395 studied by Swaters \etal would also clearly be large. We
therefore conclude that $S_2$ is a good first indicator of asymmetry
in disk kinematics as measured by Fourier decomposition.

\subsection{Conclusions}

We have examined the morphological and dynamical \ion{H}{1} symmetry
properties of a sample of moderately inclined, morphologically normal
Sa galaxies. These galaxies were known previously to exhibit kinematic
peculiarities ranging from warps to extended counter--rotating disks,
and are possibly the remnants of minor mergers.

This work confirms the normal optical morphology for which this sample 
was selected, and finds significant deviations from symmetry in both
\ion{H}{1} morphology and dynamics. This result reinforces earlier findings that
normal morphology is not an indicator of normal kinematics, and
conversely, that perturbed kinematics do not necessarily manifest as
perturbed optical morphology. Thus it appears that the time required
for the optical (stellar) component to recover from the perturbative
events which generated the distinct kinematics is smaller than the
time required for the gas to recover.

We have analyzed the circumstances of warping in the \ion{H}{1} disks
following the discussion of Briggs (1990), in which it is found that
simple warps exhibit a common structure of changing inclination within
the optical radius, followed by advancing position angle in the
direction of rotation outside of the optical radius. We find that
warps in the kinematically peculiar galaxies exhibit more complex
behavior than do the common warps seen in Briggs' sample.

The warping observed in the majority of the Sa galaxies is of 
the $15^\circ$ order in ``tip'' angle, as predicted by the numerical
simulations of minor mergers illustrated by Quinn \etal (1993) and Hernquist
\& Mihos (1995). The weak trend found in this sample between global
$n=1$ and $n=3$ asymmetry and rotation curve asymmetry $S_2$,
measuring non--circular motions, is also consistent with the
excitation of these modes due to minor mergers predicted by Hernquist
\& Mihos.  Those authors also predict the formation of a highly
non--axisymmetric potential in the final stages of the merger, as well
as morphological ``clumping'' of the gas as opposed to the smoother
distribution of stars. Such predictions are consistent with the high
values of $S_2$ and $\Delta\Gamma$ measured in this sample, as well as
the striking differences in optical versus gas morphologies.

We have also demonstrated the effectiveness of a simple rotation curve
asymmetry measure, specifically $S_2$, for identifying galaxies with
global kinematic asymmetry. This quantifier, in conjunction with the
position angle and line profile measures, are methods by which
dynamical asymmetry may be estimated quickly and thus suggest a
way to search large catalogs of rotation curves for galaxies which
exhibit lopsidedness or decoupling.

Further work will be required to determine whether these findings are
represented by the larger class of Sa galaxies. It will also be
important to consider direct comparisons between the velocity fields
of numerical simulations and observed velocity fields and their
asymmetry measures, for it is in the dynamics that minor merger
remnants are identified.

\acknowledgements
This work has been partially supported by NSF grants AST--9528860 and
AST--9900695 to
MPH, and by National Space Grant College and Fellowship Program grant
NGT--40019 to DAK. This research has made use of the NASA/IPAC
Extragalactic Database (NED) which is operated by the Jet Propulsion
Laboratory, California Institute of Technology, under contract with
the National Aeronautics and Space Administration.

\eject


\begin{references}

\reference{rotcur}Begeman, K. G. 1989. \aap, 223, 47.

\reference{bert}Bertola, F., Cinzano, P., Corsini, E. M., Pizzella,
A., Persic, M., \& Salucci, P. 1996, \apj, 458, L67

\reference{briggs}Briggs, F. H. 1990, \apj, 352, 15

\reference{buta}Buta, R., van Driel, W., Braine, J., Combes, F.,
Wakamatsu, K., Sofue, Y., \& Tomita, A. 1995. \apj, 450, 593

\reference{ciri}Ciri, R., Bettoni, D., \& Galletta, G. 1995, \nat,
375, 661

\reference{c97}Conselice, C.J. 1997, \pasp, 109, 1251.

\reference{corsini}Corsini, E. M., Pizzella, A., Funes S. J., J. G.,
Vega Beltr\'an, J. C., \& Bertola, F. 1998, \aap, 337, 80

\reference{RC3} de Vaucouleurs, G., de Vaucouleurs, A., Corwin, H.G.,
Buta, R.J., Paturel, G., \& Fouqu\'e, P., 1991. {\it Third Reference
Catalogue of Bright Galaxies\/} (RC3), University of Texas Press,
Austin

\reference{gip2000}{\it GISPY, The Groningen Image Processing
System.\/} 2000, http://thales.astro.rug.nl/ \verb|~|gipsy/ 

\reference{H79}Haynes, M. P., Giovanelli, R., \& Roberts, M. S. 1979,
\apj, 229, 83

\reference{HHMRvZ}Haynes, M. P., Hogg, D. E., Maddalena, R. J., Roberts,
M. S., \& van Zee, L. 1998, \aj, 115, 62

\reference{H2000}Haynes, M. P., Jore, K. P., Barrett, E. A., Broeils,
A. H., \& Murray, B. M. 2000, \aj, 120, 703 (Paper II)

\reference{mh}Hernquist, L. \& Mihos, C. 1995, \apj, 448, 41

\reference{J96}Jore, K. P., Broeils, A. H., \& Haynes, M. P. 1996, \aj,
112, 438 (Paper I)

\reference{KJPhD}Jore, K. P. 1997, Ph. D. thesis, Cornell
University

\reference{J2000}Jore, K. P., Haynes, M. P., Barrett, E. A., \&
Broeils, A. H. 2000, in preparation (Paper III)

\reference{KHL}Kornreich, D. A., Haynes, M. P., \& Lovelace,
R. V. E. 1998, \aj, 116, 2154 (KHL)

\reference{KHLvZ}Kornreich, D. A., Haynes, M. P., Lovelace, R. V. E.,
\& van Zee, L. 2000, \aj, 120, 139 (KHLvZ)

\reference{MK}Merrifield, M. R., \& Kuijken, K. 1994, \apj, 432, 575

\reference{q}Quinn, P. J., Hernquist, L., \& Fullagar, D. P. 1993,
\apj, 403, 74

\reference{rh94}Roberts, M. S., \& Haynes, M. P. 1994, \araa, 32, 115

\reference{RSA} Sandage, A. \& Tammann, G.A. 1987, {\it Revised Shapley Ames
Catalogue of Bright Galaxies}, Carnegie Inst. of Washington Publ. 635

\reference{SFdZ}Schoenmakers, R. H. M., Franx, M., de Zeeuw,
P. T. 1997, \mnras, 292, 349 (SFdZ)

\reference{S3vA}Swaters, R.A., Schoenmakers, R. H. M., Sancisi, R., \&
van Albada, T. S. 1999, \mnras, 304, 330

\reference{thakar}Thakar, A. R., Ryden, B. S., Jore, K. P., \&
Broeils, A. H. 1997, \apj, 479, 702

\reference{gipsy}van der Hulst, J. M., Terlouw, J. P., Begeman, K.,
Zwitser, W., \& Roelfsema, P. R. 1992, in {\it Astronomical Data Analysis
and Systems,\/} edited by D. Worall, C. Biemesderfer, \& J. Barnes, ASP
Conf. Series no. 25, 131

\reference{rz97}Zaritsky, D., \& Rix, H.-W. 1997, \apj, 477, 118

\end{references}
\end{document}